\documentclass[preprint]{elsarticle}
\usepackage[utf8]{inputenc}
\usepackage{subcaption}
\usepackage{graphicx}
\usepackage{grffile}
\usepackage{epstopdf}
\usepackage{amsmath}
\usepackage{amssymb}
\usepackage{bm}
\usepackage{hyperref}
\usepackage{xcolor}
\usepackage[autostyle]{csquotes}
\usepackage{listings}
\usepackage{todonotes}

\journal{ }

\newcommand{\sym}[1]{(\protect\includegraphics[height=6pt]{styles/#1.eps})}
\newcommand{\oneD}{$1$D }
\newcommand{\twoD}{$2$D}

\begin{document}

\begin{frontmatter}

\title{A physics-informed operator regression framework for extracting  data-driven continuum models}
\author[1]{Ravi G. Patel\corref{cor1}}
\ead{rgpatel@sandia.gov}
\author[1]{Nathaniel A. Trask}
\ead{natrask@sandia.gov}
\author[2]{Mitchell A. Wood}
\ead{mitwood@sandia.gov}
\author[1]{Eric C. Cyr}
\ead{eccyr@sandia.gov}

\cortext[cor1]{Corresponding author}
\address[1]{Sandia National Laboratories - Computational Mathematics Department}
\address[2]{Sandia National Laboratories - Computational Multiscale Department}

\begin{abstract}
The application of deep learning toward discovery of data-driven models requires careful application of inductive biases to obtain a description of physics which is both accurate and robust. We present here a framework for discovering continuum models from high fidelity molecular simulation data. Our approach applies a neural network parameterization of governing physics in modal space, allowing a characterization of differential operators while providing structure which may be used to impose biases related to symmetry, isotropy, and conservation form. We demonstrate the effectiveness of our framework for a variety of physics, including local and nonlocal diffusion processes and single and multiphase flows.
For the flow physics we demonstrate this approach leads to a learned operator that generalizes to system characteristics not included in the training sets, such as variable particle sizes, densities, and concentration. 
\end{abstract}


\end{frontmatter}

%Remove Preprint submitted to Elsevier
\makeatletter
\def\ps@pprintTitle{%
  \let\@oddhead\@empty
  \let\@evenhead\@empty
  \def\@oddfoot{\footnotesize\itshape
       \hfill\today}%
  \let\@evenfoot\@oddfoot
}
\makeatother

%%%%%%%%%%%%%%%%%%%%%%%%%%%%%%%%%%%%%

\section{Introduction}

Data-driven physics models have recently emerged in response to the surge of available science and engineering data~\cite{brunton2016discovering,rudy2017data,montans2019data}. Traditionally, equations governing dynamics of a given system have been derived by arguing from first principles, typically appealing to conservation laws and using either empirical arguments or statistical mechanics/homogenization/multiscale approaches to derive closures and constitutive relations \cite{pavliotis2008multiscale,arbogast2012mixed}. Such approaches work well when simplifying assumptions can be made, representative of near-equilibrium settings~\cite{zwanzig2001nonequilibrium}, simplified geometries~\cite{michel1999effective} or small numbers of dilute species. Recently however, applications require complex models capturing multiscale, multiphysics, and nonequilibrium processes, mandating increasingly technical frameworks for prescribing dynamics, such as the Mori-Zwanzig or GENERIC formalisms~\cite{mori1965transport,zwanzig2001nonequilibrium,li2015incorporation}. In this context, data-driven models have gained traction as a means to use the ubiquity of data to prescribe models, augmenting the traditional role of  first-principles derivations. 

Given the recent advances in machine learning (ML), there has been substantial interest in determining whether techniques from deep learning~\cite{lecun2015deep,goodfellow2016deep} may be used to prescribe data-driven models, providing machine-learned surrogates and closures to augment traditional modeling and simulation approaches~\cite{baker2019workshop}. In ML, \textit{inductive biases} describe the priors and biases imposed by a learning strategy, typically exploiting domain expertise to alleviate the required size of training data~\cite{battaglia2018relational}. In scientific machine learning (SciML) contexts, this is often referred to as \textit{physics-informed} ML, and inductive biases manifest as choices of model form that enforce preservation of physical invariants or mathematical structure~\cite{wu2018physics,raissi2019physics,ling2016machine}. The challenge of SciML is to carefully balance these inductive biases, imposing an appropriate model form to extract robust and physically realizable system dynamics without losing descriptiveness. 

In the data-driven modeling literature,
approaches assuming known dynamics with unknown material parameters or constitutive relations, are differentiated from those assuming unknown dynamics. Mature techniques exists for the former, where one obtains an inverse problem for parameter estimation that may be approached with tools such as PDE-constrained optimization~\cite{biegler2003large,haber2004inversion,epanomeritakis2008newton} or Bayesian inversion~\cite{stuart2010inverse}. In the later, it is an open question how to assume a model form which will guarantee that extracted dynamics are numerically well-posed. Many techniques exist toward this end. However, the spectrum of choices can be characterized on one end by ``black box'' ML techniques, relying on the universal approximation property of neural networks to impart no explicit bias on extracted dynamics~\cite{lu2019deeponet}, and dictionary-based techniques on the other, which restrict dynamics to a sparse selection of candidate models specified \emph{a priori}~\cite{bongard2007automated,schmidt2009distilling,bright2013compressive,brunton2016discovering,rudy2017data,long2019pde}. While the former is potentially the most descriptive, the later provides more structure to explicitly impose physical biases. In this work, we extend the framework in \cite{patel2018nonlinear} for parameterizing dynamics this lies in between, allowing imposition of physical biases without imposing an explicit model form \emph{a priori} via a dictionary.

{
In this work we adopt the perspective of modal/nodal description of operators that is ubiquitous in the spectral/spectral element discretization of PDEs~\cite{boyd2001chebyshev,hesthaven2007spectral,karniadakis2013spectral}. This perspective is not particularly new in the ML literature; see e.g. \cite{patel2018nonlinear, wu2020data,zhang2020learning} for examples where this modal perspective provides an effective means of parameterizing differential operators without the need for dictionaries. In this work, we focus on developing specializations of the architectures in \cite{patel2018nonlinear, wu2020data,zhang2020learning} which allow the introduction of physical biases. We will assume that dynamics may be described by two neural networks: one parameterizing a pseudo-differential operator in modal space composed with another network parameterizing a point-wise nonlinear operator in physical space. While applicable to a general choice of spectral basis (see for example \cite{qin2019data,wu2019numerical,wu2020data,Zhang2020}), we focus here on the Fourier basis in rectangular domains with naturally imposed boundary conditions. This provides access to both fast projection operators via the Fourier transform, and an effective parameterization of differential operators via the symbol of the operator in Fourier space. In contrast to the architecture considered for example in \cite{wu2020data}, this treatment of differential operators allows handling a single scalar function of the wavevector, rather than learning a high-dimensional black-box mode-to-mode mapping, and supports the imposition of biases, including conservation form and isotropy, which may be easily imparted to extract accurate and stable data-driven dynamics. We present this approach through the perspective of \textit{operator regression}, in which one may learn an operator via observations of its action on a representative family of functions. We establish that this perspective may be interpreted as a reduced space PDE-constrained optimization problem \cite{nocedal2006numerical,hicken2013comparison} that may be implemented entirely in popular machine learning packages such as Tensorflow and pyTorch \cite{tensorflow2015-whitepaper,NEURIPS2019_9015}, allowing use of ``off-the-shelf'' neural network architectures. Our implementation is available at \url{github.com/rgp62/MOR-Physics}. 

While numerous works have sought to learn dynamics, our focus is on the interplay of inductive biases and data. While we present many necessary components under the umbrella of operator regression (the modal/nodal perspective, learning updated operators, imposing physical biases/etc.), several of these ingredients have been previously pursued within the literature. The key contribution of the current work is to carefully design the loss and architectures so that one may extract a model which is both accurate and stable for long time integration. While several works have considered learning dynamics for simple scalar problems with synthetic data, to our best knowledge there are few frameworks which can handle a general ``black-box'' form for the governing equations for real engineering data. We have gathered a collection of benchmark systems for which the necessity of physical biases becomes increasingly important. We initially start with synthetic data corresponding to simple scalar PDE, which is easily handled in many frameworks. We then progress to extraction of continuum models from realistic simulations of particle/molecular dynamics. The corresponding data sets incorporate multiscale effects and are inherently noisy, and we informally note that without imposing physical biases it is difficult (if not impossible) to extract a stable and accurate solution. We additionally consider necessary extensions to handle systems of PDEs for which the choice of architecture and loss are critical to obtain models which scale reasonably with model complexity. This contribution is critical to bridging the gap between proof-of-concept calculations for synthetic scalar PDE data (e.g. \cite{patel2018nonlinear, wu2020data,zhang2020learning}) and the model complexity necessary for large systems of interest to the engineering community consisting of multiscale, multidimensional, and multiphysics data. 

We organize this work by first introducing a general abstract framework for data-driven modeling, and then use manufactured data to highlight its attractive properties in the absence of measurement noise. In Section~\ref{sec:diffusion} we demonstrate how anomalous diffusion processes involving nonlocal operators may be extracted from particle simulations of Levy flights. In Section~\ref{sec:lammps} we consider microchannels containing both single phase and two phase Leonnard-Jonesian fluids, extracting multiscale continuum models for the coarse-grained multiphase system that generalize to particle densities, concentrations, and particle size ratios unseen in the training set. For all applications, we apply a subset of the biases discussed in Section~\ref{sec:bias}, providing particular specializations of the general framework necessary for each application. As a whole, these exemplar multiscale problems present a series of tutorials demonstrating how the general theory and physical biases may be combined for a broad range of applications.

}

%%%%%%%%%%%%%%%%%%%%%%%%%%%%%%%%%%%%%

\section{Abstract data-driven modeling framework}

We present a general framework for data-driven modeling that we will later specialize toward specific problems. As input to the framework, we consider samples of field data

\begin{equation}\label{trainingData}
    X = \{\bm{u}^i_m\}^{i=1,\hdots,I}_{m=1,\hdots,M},
\end{equation}
where $\bm{u}^i_m$  belong to a Banach space $V$  defined on a compact domain $\Omega \subset \mathbb{R}^d$. Each index $i$ refers to a time series of functions with $I$ distinct initial conditions, while the index $m$ denotes evaluation at a time point $t_m \in \mathbb{R}_+$ with $t_0=0$. We further assume there exists a function $\bm{u}(\bm{x},t) \in \mathbf{V}\times \mathbb{R}_+$ that satisfies dynamics of the form
\begin{equation}\label{eqn:dynamics}
\begin{aligned}
 \partial_t \bm{u} & = \mathcal{N}\bm{u}, \\
     \mathcal{B}\bm{u} & = 0, \\ 
      \bm{u}(\bm{x},0) & =\bm{u}_0^i(\bm{x}),
\end{aligned}
\end{equation}
where $\mathcal{N}$ is a spatial operator, $\mathcal{B}$ is a known boundary operator, and $\bm{u}^i_0$ are known initial conditions in $X$, as described above. 

We seek to determine the dynamics $\mathcal{N}$ so that $\bm{u}(\bm{x},t_m)$ is close to $\bm{u}_m$ for all $m$ in an appropriate norm. While we consider in this work a first-order time derivative appropriate for conservation laws, the process generalizes naturally to higher-order time derivatives.

\subsection{Integral Evolution Operator}
By integrating the principal component of Eq.~\ref{eqn:dynamics} in time, we obtain the update operator $\hat{\mathcal{L}}_{t_2-t_1}$ prescribing the increment of the dynamical system state over a finite interval
\begin{equation}
    \bm{u}_2 - \bm{u}_1 = \int^{t_2}_{t_1} \mathcal{N}\bm{u}\,dt := \hat{\mathcal{L}}_{t_2-t_1}\bm{u}
\end{equation}
where $\bm{u}_i$ is the solution at $t_i$.
By restricting to a uniform time mesh (for all $n$, $t_{n+1}-t_n = \Delta t$), assuming dynamics are autonomous (i.e. $\mathcal{N}$ does not depend on time), and applying one-point quadrature, we obtain for all $n$ the approximate update operator 
\begin{equation}
    \bm{u}_{m+1} - \bm{u}_m = \mathcal{L}_{\Delta t}\bm{u}_{m+\frac12} \approx  \hat{\mathcal{L}}_{t_{m+1}-t_{m}} \bm{u} .
\end{equation}
where $\mathcal{L}_{\Delta t}[\bm{u}_{m+\frac12}] = \Delta t \mathcal{N}\bm{u}_{m+\frac12}$. The evaluation point $\bm{u}_{m+\frac12}$ denotes the choice of one-point quadrature scheme; selecting $\bm{u}_{m+\frac12}$ as $\bm{u}_{m}$ or $\bm{u}_{m+1}$ yields regression of explicit Euler or implicit Euler dynamics, respectively. For simplicity in this work, we will consider explicit dynamics and drop the $\frac12$ subscript, adopting $\mathcal{L}_{\Delta t}=\mathcal{L}$. Note that in the traditional forward modeling context, this encapsulates both explicit Euler integrators and multi-stage integrators such as Runge-Kutta; in either case only the state at the previous timestep is used to advance in time.

To regress the dynamics of Equation \ref{eqn:dynamics}, we introduce a parameterized family of operators $\mathcal{L}_\xi$ (here the subscript indicates the parametrization while the time step $\Delta t$ is assumed)  and use the data in Equation \ref{trainingData} to estimate appropriate parameters $\xi$ such that
\begin{equation}\label{eqn:discreteDynamics0}
    \bm{u}_{m+1}-\bm{u}_m = \mathcal{L}_\xi \bm{u}_{m},
\end{equation}
or equivalently, that data at $t^{n+1}$ may be related to the initial condition via repeated application of the update operator
\begin{equation}\label{eqn:discreteDynamics}
    \bm{u}_{m+1} = \left(\mathcal{I} + \mathcal{L}_\xi\right)^{m+1} \bm{u}_0,
\end{equation}
where $\mathcal{I}$ denotes the identity map.

\subsection{Operator Regression}\label{operatorRegression}
To define the optimal parameterization $\xi$, we introduce the following \textit{abstract operator regression problem}. Consider a Banach space $V$, with finite dimensional subspace $U$. Assume one has training data of the form 
$$X = \left\{f_n,{\mathcal{L}}[f_n]\right\}_{n=1}^N,$$
where $f_n$ are sampled from $U$ and $\mathcal{L}:U\rightarrow V$. Then, introducing a parameterized family of operators $\mathcal{L}_\xi:U\rightarrow V$, we seek a choice of $\xi$ such that $\mathcal{L}_\xi$ provides an optimal characterization of $\mathcal{L}$ in the following sense.

\begin{equation}\label{abstractRegression}
    \xi^* = \underset{\xi}{\textrm{argmin}} \sum_{f \in U} ||{\mathcal{L}}[f]-\mathcal{L}_\xi[f]||^2,
\end{equation}
where $||\cdot||$ denotes an appropriate norm over $V$.

While we pursue this problem in the context of data-driven modeling, it may be studied in a purely mathematical context as a generalization of classical regression. In regression, one traditionally seeks a characterization of a function $f$ at input points $x \in \Omega$, while in operator regression we instead seek a characterization of the operator $\mathcal{L}$ at input functions $f \in V$. 

To perform data-driven regression of the dynamics in Equation \ref{eqn:dynamics}, we specialize Equation \ref{abstractRegression} and find a choice of $\xi$ which best matches the data (Equation $\ref{trainingData}$). This corresponds to selecting an appropriate space $U$ such that $X$ characterizes the range of $\mathcal{L}$. In this work we select as norm the mean-square norm evaluating the action of the operator at a collection of points $Q = {\left\{x_q\right\}_{q=1}^{N_Q}} \subset \Omega$.

\begin{equation}\label{abstractRegression2}
    \xi^* = \underset{\xi}{\textrm{argmin}} \sum_{\substack{\bm{u}^i_0 \in X_{m=0}\\ x_q \in Q }} |\bm{u}^i_m (x_q) - \left(\mathcal{I}+\mathcal{L}_\xi\right)^m \bm{u}^i_0(x_q)|^2.
\end{equation}

To apply this framework, one must specify the following choices:
\begin{center}
\begin{enumerate}
    \item Choice of reconstruction space $U$
    \item Choice of parameterization for $\mathcal{L}_\xi$
\end{enumerate}
\end{center}
We explore both choices in the following sections \ref{choosingV} and \ref{choosingL}. To motivate the choice of $U$, we first relate this perspective to traditional PDE-constrained optimization.

An equivalent PDE-constrained optimization formulation
of operator regression is defined as the equality
constrained quadratic program:
\begin{equation}\label{pdeconstraints}
   \begin{aligned}
        &\underset{\mathcal{N}}{\textrm{argmin}} & 
         \sum_{i=1}^I ||\bm{v}^i(\bm{x}_m,t_m)-\bm{u}^i_m||^2 \\
        &\textrm{subject to} \  &\bm{v}^i(\bm{x},t_n) - \bm{u}^i_0(\bm{x})= \int_0^{t_n} \mathcal{N} \bm{v}^i\, dt, \\
        &                           
        &\mathcal{B}\bm{v}^i = 0, \\ 
        &                           &\bm{v}^i(\bm{x},0)=\bm{u}^i_0(\bm{x}),
    \end{aligned}
\end{equation}
where $\mathcal{B}$ and $\bm{u}^i_0$ are boundary and initial data. The constraint is written in integral form, as opposed to the form in Eq.~\ref{eqn:dynamics}, to match the form of the increment operator.
Computing the gradient of the objective function subject to the equality constraint typically requires calculation of forward and adjoint problems.
The adjoint computation is often burdensome to implement and not always available within a PDE solver, and it is therefore natural to pursue the unconstrained operator regression formulation defined in Eq.~\ref{abstractRegression2}.
As we will discuss in $\ref{choosingL}$, we will pursue neural network parameterization of operators. One convenient consequence of this choice is that the machine learning frameworks Tensorflow~\cite{tensorflow2015-whitepaper} and pyTorch~\cite{NEURIPS2019_9015} enable automatic computation of the derivatives necessary to handle Eq.~\ref{abstractRegression2}. 

For a certain choice of $U$ and $\mathcal{L}$, the operator regression may be interpreted as a reduced space \cite{nocedal2006numerical,hicken2013comparison} approach to Equations \ref{pdeconstraints}. We first note that due to the assumption of uniform discretization in time, the first constraint is automatically satisfied. This follows trivially from 
$$\bm{v}^i(\bm{x},t_n) - \bm{u}^i_0(\bm{x})= \int_0^{t_n} \mathcal{N} \bm{v}^i\, dt = \left(\mathcal{I} + \mathcal{L}_\xi\right)^{n+1}
\bm{u}^i_0.$$
To satisfy the second constraint, one must choose the space $U\subset V$ to be boundary-conforming, i.e. for any $f \in U$ we have $\mathcal{B}f = 0$. From this perspective, our operator regression framework is equivalent to a PDE-constrained optimization approach which ``plays well'' with machine learning libraries implementing unconstrained optimizers, allowing one to explore parameterizations $\mathcal{L}_\xi$ that use ``off-the-shelf'' neural network architectures without the need to develop interfaces to PDE-constrained optimization libraries.

\subsection{Choice of reconstruction space}\label{choosingV}
The choice of $U$ must enforce the boundary operator constraint (constraint two in Equation \ref{pdeconstraints}) while supporting a broad class of parameterized operators $\mathcal{L}_\xi$. In principle any boundary-conforming basis satisfies the first requirement. In light of the second, we consider modal spaces typical of spectral methods for PDEs, such as Fourier series or orthogonal polynomials. These bases generally provide a natural description of differential operators in \textit{modal space} and nonlinear operators in \textit{physical space}, which we explain below.

Specifically, we seek $U$ supporting description of differential operators via an invertible mapping $\Pi:U \rightarrow \mathbb{R}^{dim(U)}$ and a bounded operator $S$, so that
\begin{equation}\label{linearDiff}
    D^\alpha = \Pi^{-1} \circ S \circ \Pi.
\end{equation}
As a particular example, defining $U$ as the range of a truncated Fourier series will provide two desirable features. First, any differential operator may be parameterized via the Fourier symbol of $D^\alpha$ by selecting $S(\kappa) = \sum_\alpha C_\alpha \kappa^\alpha$, for multi-index $\alpha$ and coefficients $C_\alpha$, where $\kappa$ denotes wave-number. Secondly, $\Pi$ may be selected as the Fourier transform to allow a fast algorithm. For these reasons we will consider the Fourier basis in this work, however it is natural to consider whether one may select other choices for $U$ more appropriate for complex geometries. 

We gather the relevant conditions on $U$ characterizing broader choices of modal/physical spaces. Consider a complete basis such that $\forall f \in U$, $f(\bm{x}) = \mathbf{c}^\intercal \mathbf{P}(\bm{x})$, where $\mathbf{c}$ and  $\mathbf{P}$ are vectors of coefficients and basis functions, respectively. We assume the function space is characterized via a collection $\Lambda$ of degrees of freedom (DOFs), and that these DOFs are unisolvent over $U$ in the sense that there exists a unique bijection mapping $\mathbf{c}$ to $\Lambda$. In the modal setting, operators are described in terms of $\mathbf{c}$; in the physical, operators are described in terms of $\Lambda$. In the physical, nonlinear operators are often more easily described; e.g. for nodal DOFs one may sparsely evaluate the operator $u^2$ by simply squaring the DOFs at each node, whereas a modal description may yield a dense operator. Following from unisolvency, there exists an invertible mapping $\Pi:\Lambda \rightarrow \mathbf{c}$. As an example, the choice of orthogonal polynomials rather than Fourier series for $U$ fits the desired form of Equation \ref{linearDiff}, but would instead provide a differentiation matrix for $S$ and $L^2$-projection for $\Pi$. Recent work \cite{qin2019data,wu2019numerical,wu2020data} adopts this viewpoint, which may be appropriate for generalizing the current approach to complex geometries via spectral element bases.

Regarding imposition of boundary conditions, in this paper we will assume data to be arbitrarily smooth $V = C^\infty$, and denote the space of Fourier series $U = \left\{ \sum_{j=0}^J c_j \exp \left( i j 2 \pi x\right) \right\}$, with the following Dirichlet and Neumann boundary-conforming restrictions: 
$$U_D = \left\{f \in U \text{ such that } f(0)=f(1)=0\right\},$$
$$U_N = \left\{f \in U \text{ such that } \frac{df}{dx}(0) = \frac{df}{dx}(1) = 0 \right\}.$$ 
Note that this reduces to working with sine and cosine series. We use the type II discrete sine transform (DST) and discrete cosine transform (DCT) \cite{numrecipes} so that we may take as $Q$ in Equation \ref{abstractRegression2} the set of evenly spaced points,  $0, \hdots, J-1$, the boundary conditions are satisfied at $j=-1/2$ and $j=J-1/2$. 
For problems in two-dimensions, we consider tensor-product generalizations of these spaces with appropriate boundary conditions.

\subsection{Choice of operator parameterization}\label{choosingL}
Motivated by the form of Equation \ref{linearDiff}, we will consider in this work the following parameterization of $\mathcal{L}_\xi$.

\begin{equation}\label{candidateModal}
    \mathcal{L}_\xi = \Pi^{-1} \circ g_{\xi_1} \circ \Pi \circ h_{\xi_2}
\end{equation}
Again, $\Pi$ represents the mapping from physical to modal space, thus $\Pi$ and $\Pi^{-1}$ are the Fourier transform and its inverse, respectively. In general, $\Pi$ is the $L^2$-projection onto $U$, and $\Pi^{-1}$ is evaluation $\Pi^{-1}[u] = \mathbf{c}^\intercal \mathbf{P}(x)$. The functions $g_{\xi_1}$ and $h_{\xi_2}$ are neural networks whose architecture will be given later. Following the discussion in the previous section, $g$ and $h$ are designed to characterize differential operators in modal space and nonlinearities in physical space, respectively. We note a slight abuse of notation, where a given $h$ may map from $U\rightarrow V$ (consider e.g. $h=u^2$, and $U = \left\{\sin(x)\right\}$, for which $h(u)\not\in U$) so that the composition $\Pi\circ h$ is not well defined. If one works with a discrete Fourier transform, then the evaluation of $h(u)$ at discrete points implicitly provides the requisite projection of $V \rightarrow U$; we assume this to be understood when we write $\Pi \circ h$.

Several works consider various parameterization for operator-regression-like tasks in the literature \cite{wu2019numerical,lu2019deeponet,trask2019gmlsnets,Bar-Sinai}. Broadly, they may be characterized as employing varying degrees of inductive biases, balancing the expressivity of parameterization against the number of parameters and consequent data required for training. In increasing order of bias: some works use only neural networks to characterize operators, imposing no explicit bias; some assume the network may be expressed as a stencil; and others assume one may draw upon domain expertise to build a dictionary of potential model forms. The form of Equation \ref{candidateModal} assumes that the dynamics may be characterized as the composition of a differential operator and a nonlinear operator, using $\Pi$ as an encoder into modal space. {In \cite{wu2020data}, the authors consider a similar encoding into modal space, but do not include a nonlinear network, and treat $g_{\xi_1}$ as a black-box mapping from modal coefficients to modal coefficients. By instead treating the modal operator as a parameterization of the Fourier symbol, we exchange the $dim(U)$-dimensional problem for a one-dimensional one. This provides a further benefit of allowing one to use the Fourier symbol as a means of imposing biases, which we explore in the following section.}

\section{Inductive bias} \label{sec:bias}

For many scientific machine learning tasks, it is infeasible to conduct the expensive experiments or fine grain simulations necessary to extract a suitable model. Models are often improved via data augmentation \cite{Dieleman2015,Wong2016,Wang2017,Inoue2018}, wherein one \textit{a priori} knows equivariances of the feature to label map and includes transformed feature, label pairs in the training set. For example, Dieleman et al.~\cite{Dieleman2015} seek a model to predict galaxy morphology from galaxy images and asserts that the prediction should be invariant with respect to rotations. The authors augment the training set with rotated versions of the images and were able to achieve $>99$\% prediction accuracy.

However data augmentation increases the difficulty of training and, particularly for continuous transformations such as rotation, it is unclear how thoroughly one should sample the transformations. Building inductive biases directly into the model alleviates these concerns by restricting the search space of models to only those \textit{a priori} deemed feasible. For instance it is common in image classification tasks to use convolutional neural networks to enforce translational invariance \cite{Lawrence1997,Krizhevsky2012}. An active area of research seeks models with built-in equivariances \cite{Meltzer2019,Thomas2018,Esteves2017,Worrall2018}.

\begin{figure}
    \centering
    \includegraphics[width=\textwidth]{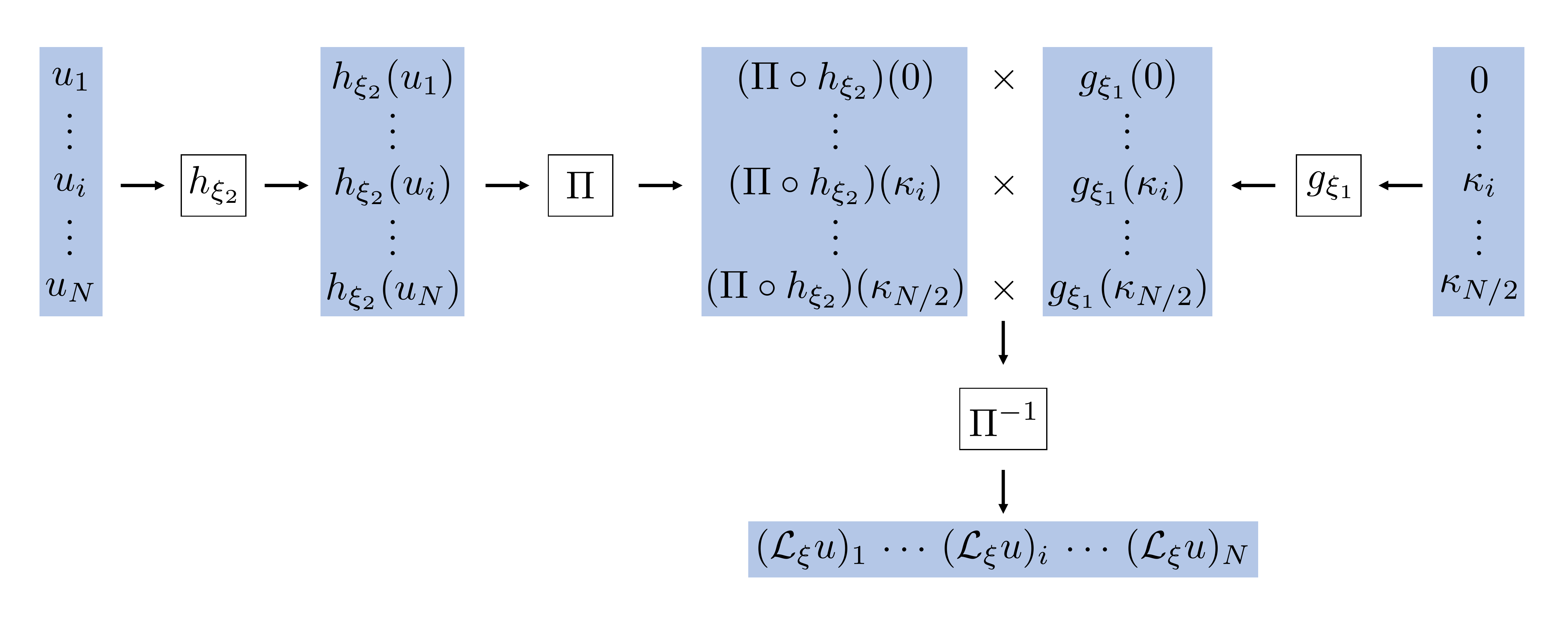}
    \caption{Application of $L_{\xi}$ to the function $u$ where $\Pi$ is the discrete Fourier transform. $u_i$ is the function evaluated at collocation point, $x_i$.}
    \label{fig:diagram}
\end{figure}

We impose inductive biases by choosing the form of Eq~\ref{candidateModal}. For regression problems using the Fourier basis, we use the following parameterization of the PDE,
\begin{equation} \label{eq:fourier_p}
    \partial_t u = \mathcal{L}_{\xi} u = \Pi^{-1} \circ g_{\xi_1}(\kappa) \Pi \circ h_{\xi_2} \circ u
\end{equation}
where $h$ is a nonlinear, point-wise operator, without explicit dependence on $x$, i.e. $(h \circ u)(x) = h(u(x))$. Figure~\ref{fig:diagram} diagrams the architecture for the discretized  $\mathcal{L}_{\xi}$. This parameterization satisfies translational invariance due to the shifting property of the Fourier transform. If $u$ satisfies Equation~\ref{eq:fourier_p}, so does a shifted function, $u(x) \rightarrow u(x - x_0) = T_{x_0}u$ because translation commutes with $\mathcal{L}_{\xi}$,
        \begin{equation}
        \begin{aligned}
            \Pi^{-1} \circ g_{\xi_1}(\kappa) \Pi \circ h_{\xi_2} \circ T_{x_0} = \Pi^{-1} \circ g_{\xi_1}(\kappa) \Pi \circ T_{x_0} \circ h_{\xi_2} \\
           =\Pi^{-1}  \circ T_{x_0} g_{\xi_1}(\kappa) \Pi \circ h_{\xi_2} = T_{x_0} \circ  \Pi^{-1} \circ g_{\xi_1}(\kappa) \Pi \circ h_{\xi_2} \\
        \end{aligned}
        \end{equation}
We consider the benefits of enforcing three additional inductive biases,

\begin{enumerate}
    \item Reflective symmetry, if $u$ satisfies Equation~\ref{eq:fourier_p}, so does $-u$. This symmetry may be enforced by letting $h(u) = \mathrm{sign} (u) \hat{h}(|u|)$, where $\hat{h}$ is a neural network. By linearity, $\Pi^{-1} \circ g(\kappa) \Pi \circ h \circ (-u) = -\Pi^{-1} \circ g(\kappa) \Pi \circ h \circ u$.
    \item Isotropy. This symmetry may be enforced by letting $g(\kappa) = \hat{g}(||\kappa||)$, where $\hat{g}$ is a neural network. By the rotation property of the Fourier transform, for a rotated function, $u(x) \rightarrow u(Rx) = Ru$,
        \begin{equation}
        \begin{aligned}
            \Pi^{-1}  \circ \hat{g}(||\kappa||) \Pi  \circ h \circ R = \Pi^{-1}  \circ \hat{g}(||\kappa||) \Pi  \circ R  \circ h \\
           =\Pi^{-1} \circ R \circ \hat{g}(||\kappa||) \Pi \circ h = R \circ \Pi^{-1} \circ \hat{g}(||\kappa||) \Pi \circ h 
        \end{aligned}
        \end{equation}
        where $(R \circ \Pi \circ v)(\kappa) = (\Pi \circ v)(R \kappa)$. This restriction requires $h$ to be point-wise operator without explicit dependence on $x$.
    \item Conservation. This constraint may be enforced globally by letting $g(\kappa) = \chi(\kappa) \hat{g}(\kappa)$, where $\hat{g}$ is a neural network and $\chi(\kappa) = 
    \left\{ \begin{matrix}
0\textrm{ for }\kappa=0 \\
1 \textrm{ otherwise}
\end{matrix} \right.$. Integrating Equation~\ref{eq:fourier_p} over space, 
        \begin{equation}
        \begin{aligned}
        \int \partial_t u dx = \int \Pi^{-1} \circ g(\kappa) \Pi  \circ h  \circ u dx \\
        = \frac{d}{dt} ( \Pi \circ u) (0) = g(0) (\Pi \circ h \circ u)(0) = 0
        \end{aligned}
        \end{equation}
        This bias can be imposed in models using the cosine basis and an equivalent parameterization as Equation~\ref{eq:fourier_p} by a similar argument.
\end{enumerate}

These inductive biases are physically motivated. For example, the response of a material is often assumed to be constant at every point in the material and in all directions, so translational invariance and isotropy are commonly supposed in mechanics models. Conservation, e.g. of mass, momentum, or energy, is often a fundamental property of a physical system. The reflective symmetry considered here is representative of symmetries encountered in many physically relevant PDEs.

%%%%%%%%%%%%%%%%%%%%%%%%%%%%%%%%%%%%%

\section{Validation for abstract operator regression problems} \label{sec:validation}

To study the effectiveness of the neural network parameterization, this section learns known steady-state spatial operators on a periodic domain. Applied to Eq.~\ref{abstractRegression}, the goal is to find a pair of parameterized operators to match two spatial operators
\begin{equation} \label{eq:spatial_problem}
{\mathcal{L}}_{\xi_1} u \sim \partial_x u^2 \mbox{ and } {\mathcal{L}}_{\xi_2} u \sim \nabla^2 u.
\end{equation}
The observations, $u$ and $\hat{\mathcal{L}}u$, are sampled using the following strategy. Each input function, $u$, is generated by a Fourier expansion,
\begin{equation}
    u = \sum_{|\kappa|<=10} c_{\kappa} \exp(j \kappa \cdot x)
\end{equation}
where $c_{-\kappa} = \bar{c}_{\kappa}$, $|c_{\kappa}|=1$, and $\arg(c_{\kappa})$ is sampled from the uniform distribution, $\mathcal{U}[0,2\pi]$. The action of the differential operator, $\hat{\mathcal{L}}u$, is defined using Equation~\ref{candidateModal}. For each problem, we generate training sets by sampling $N$ functions and computing the actions of the operator. A similar parametrization is used for both the advection operator and Laplacian,
\begin{equation} \label{eq:spatial_param}
    \mathcal{L}_\xi = \Pi^{-1} \, g_\xi(\kappa)\, \Pi \, h_\xi(u)
\end{equation}
where $g_\xi$ and $h_\xi$ are neural networks. Table~\ref{tab:spatial_param} shows the parameters used for the networks and training.

To facilitate comparisons with the spatial operators, the parameterization is modified
in a post processing step to remove redundant degrees of freedom. For instance, the
choice of $f_\xi$ and $g_\xi$ can be easily modified by a scaling to give an equivalent
operator. To this end, for both operators in Eq.~\ref{eq:spatial_problem} in post processing we shift $h_{\xi}$ and $g_{\xi} (0)$ such that $h_{\xi}(0)=0$. Additionally,
for Burger's we scale $h_\xi$ and $g_\xi$ such that $\frac{d^2 h}{du^2}=2$ at $u=0$.
For the Laplacian, we scale $h_\xi$ and $g_\xi$ such that $\frac{d h_\xi}{du}=1$ at $u=0$.

\begin{table}
\begin{center}
    \caption{Hyperparameters for spatial operator regression}
    \begin{tabular}{c|c} \label{tab:spatial_param}
        Parameter & Value \\
        \hline
        Network width & 4 \\
        Network depth & 4 \\
        Activation function & elu \\ 
        Optimizer & Adam \\
        Learning rate & 0.001 \\
        Batch size & 10 \\
        Epochs & 1 \\
        Grid size & $256$ for $\partial_xu^2$; $128 \times 128$ for $\nabla^2 u$ \\ 
        Training set size & 800,1600,3200,6400,12800
    \end{tabular}
\end{center}
\end{table}

\begin{figure}
    \centering
    \includegraphics[width=4.5in]{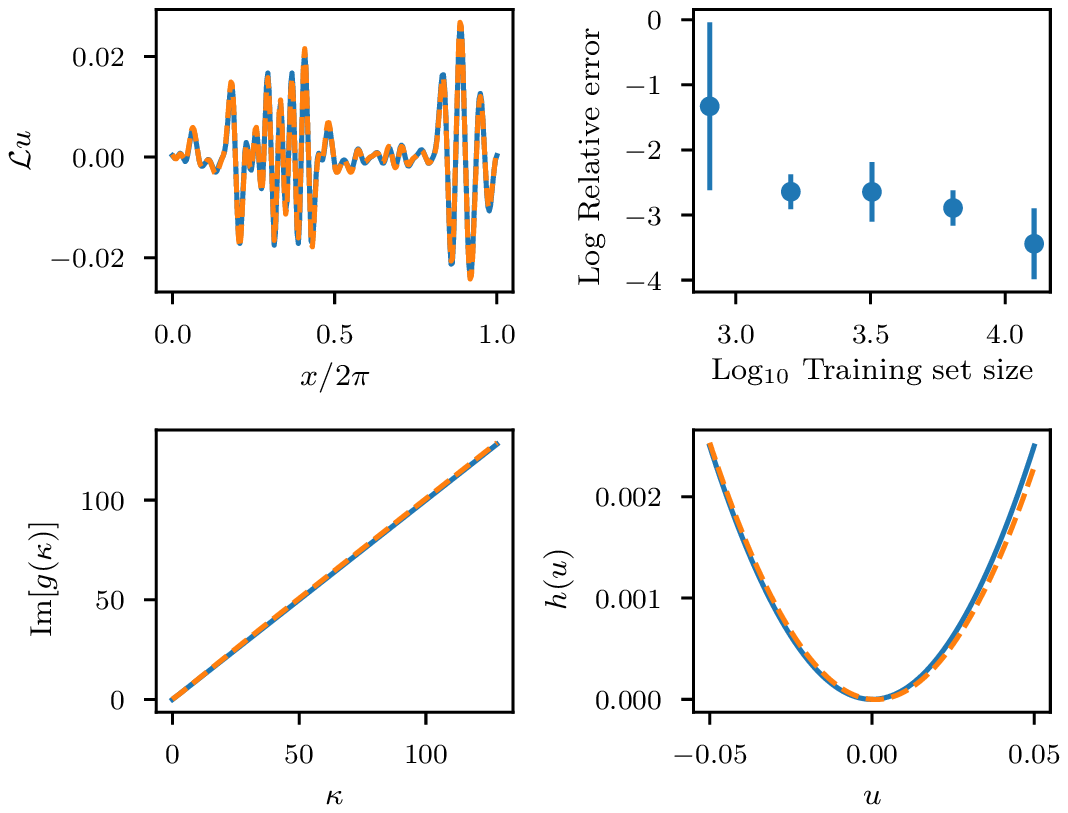}
    \caption{Application of the Burgers operator, $\partial_x u^2$, \sym{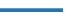} and regressed operator \sym{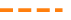} to a function (\textit{top left}). $\mathrm{Im}[g(\kappa)]$ (\textit{bottom left}) and $h(u)$ (\textit{bottom right}) for  both operators. Relative error, $||\mathcal{L}u - \partial_x u^2||/||\partial_x u^2||$ vs. training set size (\textit{top right}).}
    \label{fig:burgers}
\end{figure}

Figure~\ref{fig:burgers} shows the results of learning a \oneD nonlinear spatial operator motivated by Burger's equation: $\partial_x u^2$.  The upper left image shows a representative sample from the solution space with the true value of the differential operator in blue and the regressed evaluation in orange. The log of relative error in minimizing the loss is in the upper right with errors bars representing one standard deviation of the log error. The high accuracy of the relative loss, indicates that the learned operator is relatively accurate. This is confirmed by the lower two images, which show learned (orange) and exact (blue) modal function $g$ on the left, and the nonlinear function $h$ on the right. 

\begin{figure}
    \centering
    \includegraphics[width=4.5in]{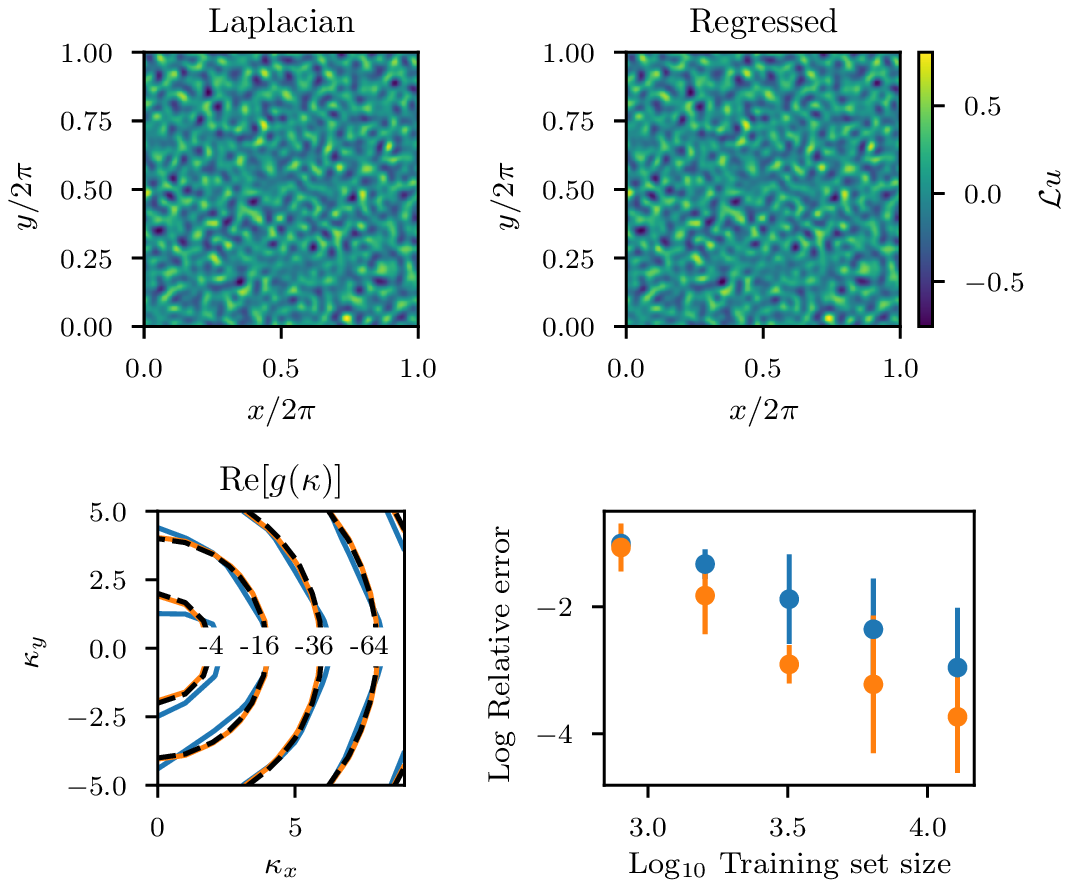}
    \caption{Application of Laplacian to a function, $\nabla^2 u$, (\textit{top left}) and application of the regressed operator with isotropy assumption to the same function, $\mathcal{L}u$, (\textit{top right}). Real part of the symbol of the Laplacian \sym{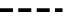}, regressed operator without isotropy assumption \sym{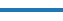}, and regressed operator with isotropy assumption \sym{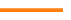} (\textit{bottom left}). Relative error, $||\mathcal{L}u - \nabla^2 u||/||\nabla^2 u||$ vs. training set size (\textit{bottom right}).}
    \label{fig:laplace}
\end{figure}

Figure~\ref{fig:laplace} similarly learns $\nabla^2 u$ in $2$ spatial dimensions. An examination of the effect of including an isotropy assumption (see Section~\ref{sec:bias}) in the construction of the parameterization is studied. Here the lower right image shows the relative error and demonstrates that a higher-level of agreements is achieved when the isotropy assumption is applied (orange) rather than ignored (blue). Again, this observation is clear from the contour plot of the value of $g$ for different values of the modal space. The parameterization enforcing isotropy agrees strongly with the exact values for the Laplacian operator (dotted black). However, the agreement with the unrestricted parameterization is not as strong.  This demonstrates the importance of enforcing appropriate inductive biases to improve the quality of the learned operator.

%%%%%%%%%%%%%%%%%%%%%%%%%%%%%%%%%%%%%

\section{Diffusion Example: Extraction of anomalous diffusion from molecular data} \label{sec:diffusion}
As a canonical example of a particle system exhibiting anomalous diffusion\cite{sokolov2005diffusion}, we consider the evolution of a number of particles described by the Levy process,
\begin{equation} \label{eq:brownian-micro}
    dX_t = dZ_t,
\end{equation}
in a periodic domain, whose increment is specified by the $\alpha$-stable distribution with skewness $0$, scale parameter $\Delta t ^{1/\alpha}$, and location parameter $0$ (i.e. $Z_{t+\Delta t} - Z_t \sim L(\alpha,0,\Delta t ^{1/\alpha},0)$). For details on the $\alpha$-stable distribution, along with algorithms and code to sample it, we refer readers to \cite{gulian2018stochastic} and references therein.

In the continuum limit, the particle density satisfies a fractional diffusion equation. For the particular choice of $\alpha = 2$, $Z_t$ one recovers the familiar Brownian motion with continuum limit corresponding to the heat equation
\begin{equation}
    \partial_t u = \Delta u,
\label{eq:brownian-continuum}
\end{equation}
while selecting $\alpha <2$ and $\alpha > 2$ describes sub-diffusion and super-diffusion processes, respectively. While in general fractional operators require careful numerical treatment~\cite{lischke2018fractional}, on periodic domains they admit a particularly simple description via a Fourier symbol with fractional exponent, which we will see fits naturally into the current framework. We will consider both one- and two-dimensional scenarios below.
\subsection{Application of abstract theory}

For operator regression the task is to \emph{train} on coarse grained data generated by Eq.~\ref{eq:brownian-micro}, and \emph{learn} a continuum representation approximating
Eq.~\ref{eq:brownian-continuum}. This provides a test that the operator regression strategy we employ can yield accurate predictions. In addition, we will demonstrate the effect that the inductive biases have on the performance of the regression. 
As a \emph{validation} experiment, a trajectory of the learned continuum operator is applied to an initial condition that was not used in training, and compared to the evolution of a corresponding molecular trajectory with identical initial condition. This will demonstrate the ability of the regressed operator to generalize to unseen initial conditions.

Due to the periodicity of the domains, the Fourier basis is used with the parameterization in Eq.~\ref{eq:fourier_p} for the spatial operator. In both \oneD and \twoD\ the continuum evolution of the density is computed by binning the particles in histograms. Regular grids are used for the histograms, as discussed in Section~\ref{sec:brownian_gen} and \ref{sec:levy_gen}, so that the transforms can be computed efficiently.

\begin{table}
\begin{center}
    \caption{Hyperparameters for diffusion operator regression}
    \begin{tabular}{c|c} \label{tab:diff_param}
        Parameter & Value \\
        \hline
        Network width & 4 \\
        Network depth & 4 \\
        $M$ & 4 \\
        Activation function & elu \\ 
        Optimizer & Adam \\
        Learning rate & 0.001 \\
        Batch size & 100 \\
        Epochs & 1 \\
        Grid size & $256$ for \oneD; $64 \times 64$ for \twoD 
    \end{tabular}
\end{center}
\end{table}

To avoid performing the optimization in Eq.~\ref{abstractRegression2} through a large number of compositions, $m$, we approximate it as,
\begin{equation}\label{reg_diff}
    \xi^* = \underset{\xi}{\textrm{argmin}} \sum_{\substack{\bm{u}_j \in X\\ x_q \in Q}} |\bm{u}_{m+j}(x_q) - \left(\mathcal{I}+\mathcal{L}_\xi\right)^m \bm{u}_j(x_q)|^2.
\end{equation}
where we iterate the above optimization problem $M$ times, setting $m=1$ for the first iteration and increasing $m$ by 1 every iteration. For the examples in this section, we use the hyperparameters listed in Table~\ref{tab:diff_param}.

{
We remark the the problem setup bears similarities to that considered in \cite{wu2020data}. An important distinction is that the update operator $\mathcal{I} + \mathcal{L}_\xi$ will be modeled by a single update, whereas \cite{wu2020data} model the update operator using a residual network. Their approach may be interpreted as learning a multi-stage update operator, at the expense of requiring a separate set of hyperparameters $\xi$ at each stage. From this perspective the current work may be considered as a special case corresponding to a single layer ResNet. Practically this amounts to a smaller number of hyperparameters, learning one operator $\mathcal{L}_\xi$ rather than multiple for each substage. When we later move to more difficult problems however, we will require a more involved loss in Eqn. \ref{reg_lammps}, and at that point we will lose the ResNet interpretation.
}
\subsection{Results}

\subsubsection{Generation of training data from \oneD Brownian motion} \label{sec:brownian_gen}
The training data is generated by particle motion evolved according to Eq.~\ref{eq:brownian-micro}. Particle positions, $X(t=0)$, are initialized by sampling from a parameterized multi-von Mises distribution,
\begin{equation}
        P(X) = \frac{1}{10} \sum_{i=1}^{10}  \frac{\exp(\kappa_i \cos(x - \mu_i))}{2 \pi I_0}
\end{equation}
where $I_0$ is the zeroth order modified Bessel function of the first kind.  For a single parameterization ($\kappa_i$ and $\mu_i $ are fixed) the initial location of each particle is generated by first drawing $i$ from a discrete uniform distribution, $\mathcal{U}[0,5]$, and then sampling from the corresponding $f_i$. The trajectories, $X(t)$, are evolved in the periodic domain $[0,2\pi)$ using the Euler-Maruyama algorithm with a timestep size of $0.001$. At each timestep, we compute an empirical particle density, $u(x,t)$ by binning the particles in 256 uniform cells. We generate $100$ such particle trajectories each containing $51200$ particles at randomly selected parameter values.  The five scale and location parameters $\kappa_i$ and $\mu_i$ are sampled from a uniform distribution, $\mathcal{U}[5,10]$, and a uniform distribution, $\mathcal{U}[0,2 \pi]$, respectively.
We perform 5 realizations of training using the same data, randomly initializing the neural network parameters.

\subsubsection{Extracting a continuum model from Brownian Motion}
\begin{figure}
    \centering
    \includegraphics[width=1.0\textwidth]{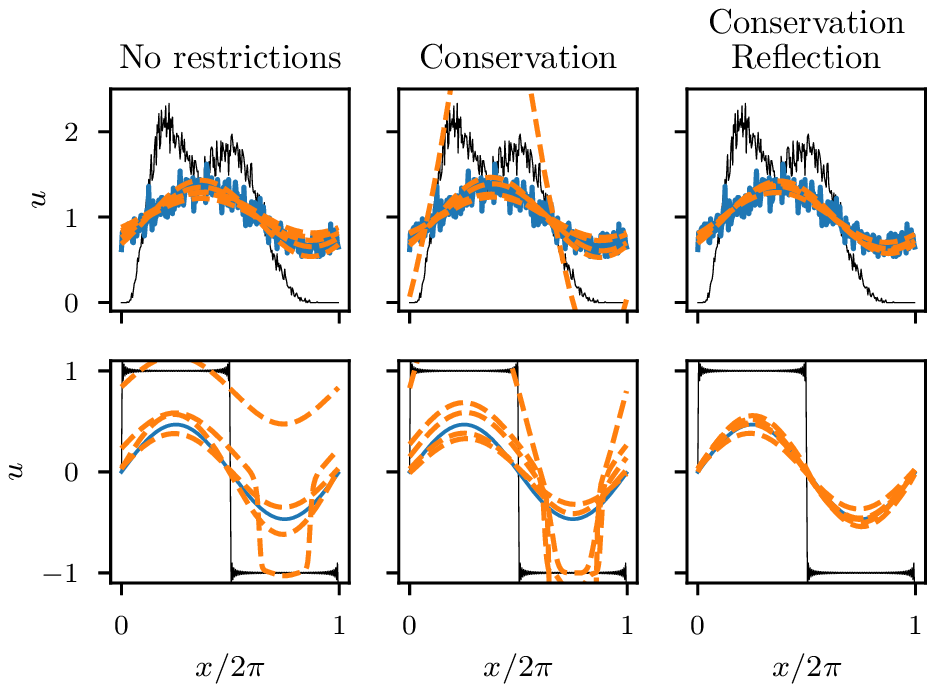}
    \caption{Evolution of the Brownian motion training data (\textit{top row}) and the heat equation validation example (\textit{bottom row}) from $t=0$ \sym{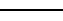} to $t=1$ \sym{C0,l,n,1} compared to evolution of $5$ realizations of the learned equation \sym{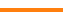}. Each column depicts the final learned solution with different physical assumptions yielding improved training and validation accuracy.}
    \label{fig:1dbrownian}
\end{figure}

The top row of Figure~\ref{fig:1dbrownian} shows the training data at the final time (in blue) compared to the final time of the trajectories evolved using
the $5$ learned operators (in orange). The black line in the images is the initial
condition. The regressed result uses different inductive biases in each column of the figure.
Although the histogram data is noisy, the Fourier operator regression with no enforced restrictions learns a spatial operator that gives a PDE with similar evolution to the particle density under Brownian motion for most of the realizations.
Although the learned equations agrees well with the training set, they fail to generalize to the validation square wave initial condition presented in the second row. In this example, a truncated Fourier series expansion of a square wave is used as the initial condition and evolved under the heat equation. The analytical solution is,
\begin{equation} \label{eq:square}
    \begin{aligned}
        &\hat{u}(x,t) = \Pi^{-1} g(\kappa) \exp(-\kappa^2 t) \\
        &g(\kappa) = \left\{ \begin{matrix}
                        0 \hfill & \mathrm{if}\ \kappa \textrm{ is even} \\
                        -512j/\pi \kappa & \mathrm{else} \hfill
                        \end{matrix}
                \right.
    \end{aligned}
\end{equation}
and shown in blue for $t=1$. The learned evolution often produce solutions that drift over time or oscillate when $u$ is negative. This situation can be rectified, however, by enforcing inductive biases on the parameterization of the learned operator.

\begin{figure}
    \centering
    \includegraphics[width=\textwidth]{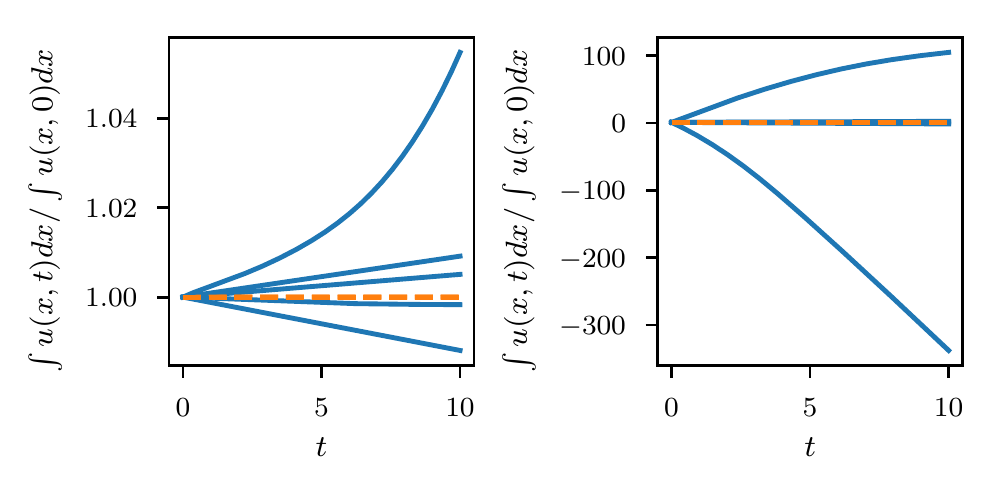} \\
    \caption{Evolution of total particle density for initial condition provided by training set (\textit{left}) and square wave initial condition (\textit{right}). Results are shown for parameterization that does not enforce any physics \sym{C0,l,n,1} and parameterization that enforces conservation \sym{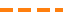}.}
    \label{fig:1dbrownian_cons_err}
\end{figure}

To explain the improvements gained from enforcing the inductive biases consider first the ``no restrictions'' validation result in Fig.~\ref{fig:1dbrownian}. Here  several of the learned approximations drift away from the exact solution.
In Figure~\ref{fig:1dbrownian_cons_err} this qualitative observation is quantitatively expressed as a time history of the
current total particle density relative to the initial particle density. For both training (left image) and validation (right image) results, the learned evolved particle density can deviate substantially from the
initial particle density. Introducing the conservation constraint, as discussed in Sec.~\ref{sec:bias}, provides a correction to the solution as observed in the middle images of Fig.~\ref{fig:1dbrownian}. These solutions don't drift; however, the validation and training solutions remain flawed.

\begin{figure}
    \centering
    \includegraphics[width=\textwidth]{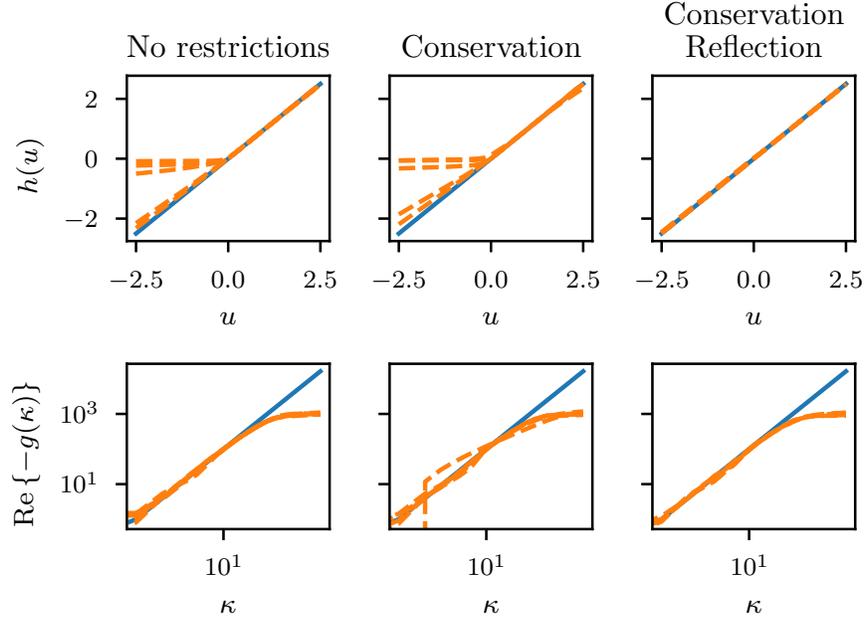}
    \caption{$h(u)$ (\textit{top row}) and $g(\kappa)$ (\textit{bottom row}) for the true heat operator \sym{C0,l,n,1} and learned operator \sym{C1,d,n,1} without enforcing physics (\textit{left column}), enforcing conservation (\textit{middle column}), and enforcing conservation and parity (\textit{right column}). }
    \label{fig:1dbrownian_cons_sym}
\end{figure}

Figure~\ref{fig:1dbrownian_cons_sym} demonstrates why the reflection constraint can be used to address these problems. The figure shows the learned and exact $h$ (top row) and $g$ (bottom row) functions. The learned $g_\xi$ corresponds well to the known value at low frequencies regardless of the enforcement of constraints. At high frequencies, the approximation deviates. This is not surprising as the spatial resolution of the original Brownian motion is bounded by the choice of histogram size. More relevant to the 
issue at hand is the function $h$. Here enforcement of the reflection constraint is critical to getting this function correct. When this constraint is unenforced, the optimizer lacks the information to fit function $h_\xi$ for $u<0$ as original training data only possesses strictly positive densities. Thus the trained solution was unable to extrapolate beyond the observed regime. When the reflection condition is included, the assumption that the behavior is independent of the sign is enforced leading to improved generalization accuracy. The end result is seen clearly in the rightmost images in Figure~\ref{fig:1dbrownian_cons_sym} where both conservation and reflection biases are enforced yielding a parameterization of the $g_\xi$ and $h_\xi$ functions that closely matches the exact functions.

\subsubsection{Generation of training data from \twoD Levy flight} \label{sec:levy_gen}
Empirical particle densities are computed in a similar manner as in Section~\ref{sec:brownian_gen}. Multi-von Mises distributions are used to generate the $x$ and $y$ coordinates of the initial particle positions,
\begin{equation}
    P(X) = \frac{1}{5} \sum_{i=1}^{5}  \frac{\exp(\kappa^d_i \cos(x - \mu^d_i))}{2 \pi I_0},
\end{equation}
where the scale factors, $\kappa^d_i$, and location parameters, $\mu^d_i$, are sampled independently for the $d$ coordinate. The particle trajectories are computed using the Euler-Maruyama algorithm with increments sampled from an isotropic stable distribution with $\alpha=1.5$ and timestep $dt=0.001$. We compute $100$ sets of $819200$ particle trajectories and compute particle densities at every timestep using histograms with $64 \times 64$ bins. As in Section~\ref{sec:brownian_gen}, $\mu_i^d$ is sampled from a uniform distribution, $\mathcal{U}[0,2 \pi]$.

We vary the distributions of $\kappa^d_i$ to generate anisotropicly biased data. For example, if $\kappa^x_i = 0$, the initial particle density will be uniform in $x$ and will remain zero for all time. In this situation, the empirical time evolution of particle density lacks information about dynamics in the $x$ direction. In the following section, we quantify the effects that anisotropicly biased data has on training. We introduce $\beta$ as the ``degree of anisotropy'' in the training set, sampling $\hat{\kappa}^x_i$ and $\kappa^y_i$ from $\mathcal{U}[1,10]$, and letting $\kappa^x_i=(1-\beta) \hat{\kappa}^x_i$.
We perform 10 realizations of training using the same data, randomly initializing the neural network parameters.

\subsubsection{Levy Flights}
The previous examples demonstrated the ability of the operator regression approach to learn integral order spatial operators, and the effects that assuming an inductive bias can have on the results. In this section, we demonstrate the advantage of using a spectral discretization of the operator to realize fractional spatial operators. In addition, we study the accuracy of the operator regression with and without the iostropic bias. This confirms, for this case, the intuitive expectation that more accuracy can be obtained with less data if the appropriate physics is included in the parameterization of the operator.

\begin{figure}[htpb]
    \centering
    \includegraphics[width=4.5in]{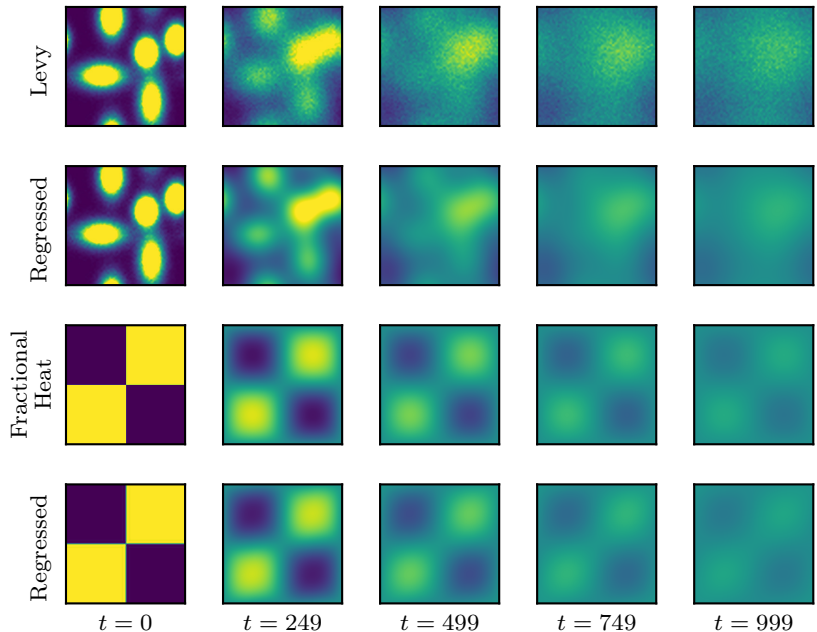}
    \caption{ Histograms showing evolution of Levy flight \textit{(first row)}. Evolution of learned operator using histogram of Brownian motion at $t=0$ as the initial condition \textit{(second row)}. Evolution of the fractional heat equation with square wave initial condition \textit{(third row)}. Evolution of learned operator on square wave initial condition \textit{(fourth row)} }
    \label{fig:2d_levy}
\end{figure}

The top row of Figure~\ref{fig:2d_levy} presents the training data generated by a Levy flight yielding a continuous fractional diffusion operator with exponent of $\alpha/2=0.75$. The second row demonstrates that the learned operator can recreate the training data. The second set of rows includes the validation result. The third row shows the time evolution of a continuous fractional heat equation. We show the analytical solution for a truncated Fourier series expansion of a square wave initial condition, 
\begin{equation} \label{eq:square_2d}
    \begin{aligned}
        &\hat{u}(x,t) = \Pi^{-1} g(\kappa) \exp(- |\kappa|^2 t) \\
        &g(\kappa) = \left\{ \begin{matrix}
                0 \hfill & \mathrm{if}\ \kappa_x \textrm{ or }\kappa_y \textrm{ is even} \\
                        -16384j/\pi^2 \kappa_x \kappa_y & \mathrm{else} \hfill
                        \end{matrix}
                \right.
    \end{aligned}
\end{equation}
This solution matches closely to the evolution of the field using the learned operator, as shown in the fourth row. 

\begin{figure}[htpb]
    \centering
    \includegraphics[width=4.5in]{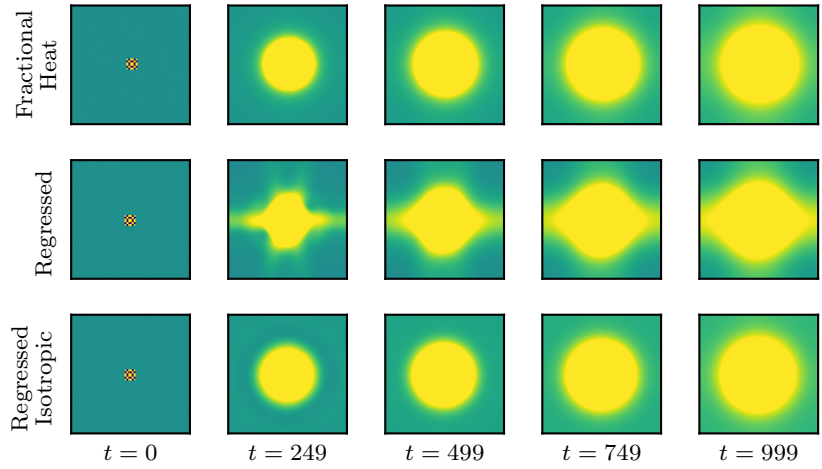}
    \caption{ Evolution of the fractional heat equation with Dirac delta initial condition \textit{(first row)}. Evolution of the learned equation without isotropy assumption with Dirac delta initial condition \textit{(second row)}. Evolution of the learned equation with isotropy assumption with Dirac delta initial condition \textit{(third row)}.}
    \label{fig:2d_levy_iso}
\end{figure}

\begin{figure}[htpb]
    \centering
    \includegraphics[width=2in]{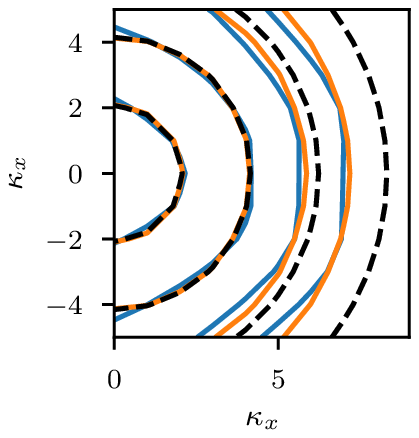}
    \caption{Real part of $g(\kappa)$ for fractional Laplacian \sym{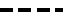}, regressed operator without isotropic assumption  \sym{C0,l,n,1}, and regressed operator with isotropic assumption   \sym{C1,l,n,1}.}
    \label{fig:2d_levy_iso_gk}
\end{figure}

To examine the impact of inductive biases, the fractional isotropic diffusion of a truncated Fourier expansion of the Dirac~$\delta$ function is considered. Figure~\ref{fig:2d_levy_iso} shows analytical solution,
\begin{equation}
    \hat{u}(x,t) = 4096 \Pi^{-1} \exp(-|\kappa|^{1.5}t)
\end{equation}
in the top row. The remaining rows show the learned diffusion of the operator without an isotropic inductive bias (middle row), and with an isotropic inductive bias (bottom row). These image clearly demonstrate that including the inductive biases helps maintain symmetry. Figure~\ref{fig:2d_levy_iso_gk}  shows isocontours of the learned spectrum. For smaller modes (with larger wave lengths) the isotropic assumption clearly benefits the learning problem. As the wavelengths shrink, however, we see that all learned operators deviate from the theory as noise overtakes signal in the training data.

\begin{figure}[htpb]
    \centering
    \includegraphics[width=4in]{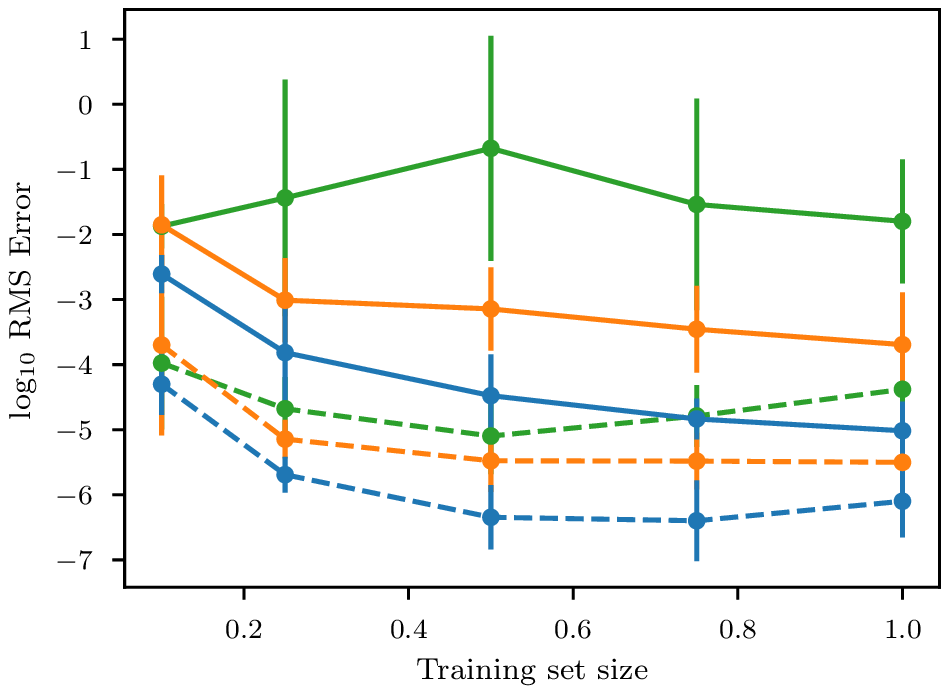}
    \caption{Test error vs. size of training set for anisotropic model trained on data with degree of anisotropy, $\beta=0.0$ \sym{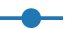}, $\beta=0.1$ \sym{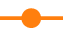}, and $\beta=1.0$ \sym{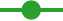}; and for isotropic model trained on data with $\beta=0.0$ \sym{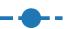}, $\beta=0.1$ \sym{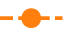}, and $\beta=1.0$ \sym{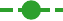}. (statistically isotropic: $\beta=0$; statistically \oneD: $\beta=1$)}
    \label{fig:2d_levy_iso_beta}
\end{figure}

To quantify the impact of including the isotropic assumption, Figure~\ref{fig:2d_levy_iso_beta} presents the error achieved in reproducing the training data using the learned operator. In this experiment, an isotropic Levy flight is used. However, the degree of anisotropy in the initial condition, specified by $\beta$ in the figure, varies over $3$ training sets. Further, the sizes of the training sets are varied. In the images, the RMS error in the learned operator is plotted as a function of the training set size, with each line representing a different degree of anisotropy. The regression used for the solid lines has no inductive bias, while the regression used for the dotted lines assumes isotropy. Comparing these two operators, it is clear that the including the inductive bias reduces the error by multiple orders of magnitude in this case regardless of the degree of anisotropy in the training data. Furthermore, this is not simply an effect of the amount of training data as the isotropic case is able to do better for all values of $\beta$. In the case of a $\beta=1$, this is a $1$d distribution of the initial condition and despite this behavior the isotropic case still performs better although it does not have the marked improvement with training set size. 

%%%%%%%%%%%%%%%%%%%%%%%%%%%%%%%%%%%%%

\section{Continuum mechanics: extraction of continuum mechanics models from molecular dynamics simulations} \label{sec:lammps}

A common problem in computational science and engineering is bridging micro- and macro- scale descriptions of a physical system. For many systems a first principles micro-scale description is available, but the range of length- and time-scales needed to simulate a complete system are outside the reach of even leadership computing facilities. To simulate such systems, accurate macro-scale models are needed.
We demonstrate the present method as a strategy for extracting continuum scale models from data collected in molecular dynamics simulations.

We consider two pressure driven, planar channel systems, a Lennard-Jones (LJ) fluid and a colloidal fluid. 
For the LJ system, as training we vary the density while for the colloidal system, we vary both the colloid concentration and the particle sizes. Our goal is to find models that accurately predict the evolution of a set of macroscopic states. We will demonstrate the scheme's ability to generalize learned dynamics beyond the concentrations, particle sizes, and densities contained in the training data.

\subsection{Application of abstract theory}

It is difficult to know which fields carry sufficient information for prediction to be accurate or even possible. Furthermore, increasing the number of fields to evolve may increase the difficulty of training. Therefore, we compare models that evolve different fields for the LJ and colloidal systems. For the LJ system, we consider a one field model that evolves only the particle momentum, a two field model that evolves the momentum and the particle density, and a three field model that evolves momentum, density, and the density weighted temperature. For the colloidal problem, we consider a two field model that evolves the large particle density and total particle momentum, a four field model that evolves the particle density and momentum for the large and small particles separately, and a six field model that evolves density, momentum, density weighted temperature for the two phases. The geometry of the systems, ensures they are statistically \oneD in the wall normal direction. Thus, we seek \oneD models for the evolution of these fields. We assume homogeneous Neumann boundary conditions for the density and temperature fields and homogeneous Dirichlet boundary conditions for the momentum fields. 

All models considered have the following parameterization,
\begin{equation}
    \begin{aligned}
    &\partial_t u^i_{{N}} = \sum^K_k \mathcal{C}^{-1} g_j^i(\kappa;\gamma) \mathcal{C} h_j^i({\bm{u}};\gamma) \\ 
    &\partial_t u^i_{{D}} = \sum^K_k \mathcal{S}^{-1} g_j^i(\kappa;\gamma) \mathcal{S} h_j^i({\bm{u}};\gamma)
    \end{aligned}
\end{equation}
where $\mathcal{C}$ and $\mathcal{S}$ are the DCT and DST discussed in Section~\ref{choosingV}, $\bm{u}$ are all the fields being evolved, $\bm{u}_{N}$ are the fields satisfying Neumann boundary conditions, $\bm{u}_{D}$ are the fields satisfying Dirichlet boundary conditions, $\gamma$ are the varying simulation parameters, and $K$ is a hyperparameter.

For the LJ system, $\gamma=\rho^*$ is the LJ reduced density and
\begin{align}
    &\textrm{1 field:\quad}(\bm{u}_{{N}},\bm{u}_{{D}}) = ([\emptyset],[p]), \\
    &\textrm{2 field:\quad}(\bm{u}_{{N}},\bm{u}_{{D}}) = ([\rho],[p]), \\
    &\textrm{3 field:\quad}(\bm{u}_{{N}},\bm{u}_{{D}}) = ([\rho,\rho T],[p]),
\end{align}
where $p$ is the particle momentum, $\rho$ is the particle density, and $\rho T$ is the density weighted temperature. For the colloidal system $\gamma = (c,d)$ is the colloid concentration and colloid particle size and
\begin{align}
    &\textrm{2 field:\quad} (\bm{u}_{{N}},\bm{u}_{{D}}) = ([\rho^L],[p^L+p^S]), \\
    &\textrm{4 field:\quad} (\bm{u}_{{N}},\bm{u}_{{D}}) = ([\rho^L,\rho^S],[p^L,p^S]), \\
    &\textrm{6 field:\quad} (\bm{u}_{{N}},\bm{u}_{{D}}) = ([\rho^L,\rho^S,(\rho T)^L,(\rho T)^S],[p^L,p^S]), 
\end{align}
where $\rho$, $p$, and $(\rho T)$ are again the density, momentum and density weighted temperature, and the superscripts, $L$ and $S$ refer to the colloid particles and solvent particles, respectively.

Using these parameterizations, we attempt to find models that capture both the transient and steady state behaviors. We found the following loss function to yield successful models,
\begin{equation}\label{reg_lammps}
    \xi^* = \underset{\xi}{\textrm{argmin}} \sum_{\substack{\bm{u}_j \in X\\ x_q \in Q \\ m \in [1,\hdots ,M]}} \frac{1}{m^\beta} |\bm{u}_{m+j}(x_q) - \left(\mathcal{I}+\mathcal{L}_\xi\right)^m \bm{u}_j(x_q)|^2.
\end{equation}
where $\beta$ is an additional hyperparameter and $m$ is sampled from a uniform distribution $\mathcal{U}[0,M]$. For $\beta>0$, this loss function preferentially penalizes update operators that fail to capture dynamics on fast timescales. Compared to the loss function in Equation~\ref{reg_diff}, this loss function does not lead to the optimizer forgetting short time scale dynamics as it learns longer time scale dynamics.

Due to the configuration of these systems, the number of particles in the domain remains constant over time, and therefore, we may assume conservation of $\rho$. In Section~\ref{sec:model} we consider the benefits of enforcing conservation on $\rho$.

\subsection{Generation of training data}
The LAMMPS particle simulation engine is used to generate the training data needed. LAMMPS evolves the dynamics of each particle based on a Verlet time integration of Newton's equations of motion, e.g. $m_i\partial^2_t r_{i} = \nabla U\{r_{ij}\}$~\cite{plimpton1993fast, tuckerman1992reversible, parrinello1981polymorphic}.
Inter-particle forces ($\nabla U$) are used in a discrete time integration algorithm to update velocities and positions. Forces for particle $i$ are accumulated by summing over all neighboring particles $j$ using the distance $r_{ij}$ between particles. 
There are many ways to select $U\{r_{ij}\}$ functionals, for instance recent work has applied machine learning techniques to achieve a significant advances in model fidelity~\cite{zuo2020performance}. However, for this work we have employed the colloid potential energy functional of Everaers \textit{et. al.}~\cite{everaers2003interaction} that reduces to a Lennard-Jones model for point masses.
Simulations of a two-phase mixture of large and small particles are initialized on a regular face-centered cubic lattice with particle types assigned at random in a domain with periodic boundary conditions. This initial spatial ordering evolves quickly ($<10^3$ timesteps) to approach steady-state flow.
For the Lennard-Jones system, the volume of the cell is varied to produce particle densities, $\rho^* \in [0.005,0.01,0.02,0.05,0.1,0.2,0.5,1]$ to modulate the training space. For the colloidal system, the concentration of large particles is varied as, $c \in [0.2, 0.15,0.1,0.05,0.025]$, the colloid particle radius is varied as, $d \in [2,2.5,3,4,5]$,
and the energy pre-factor terms set to 10:1 with respect to the larger particles. The Lennard--Jones fluid in both systems is constructed by setting the particle radius to zero.
All energies, times and distances are provided in reduced LJ units.

To satisfy the Dirichlet BC, a rectangular channel is constructed by constraining the initialized particles at the $\pm$X boundary to be fixed in time. This creates a rigid wall that prevents particles from passing through the boundary. 
A uniform particle flow is simulated by adding a constant force (in addition to $\nabla U\{r_{ij}\}$) to the remaining mobile particles in the $+$Y direction. A constant temperature is maintained by a Nose-Hoover thermostat.
This results in steady-state dynamics where the number of particles (of each type), simulation volume and temperature are conserved quantities (i.e. the canonical ensemble).
Training data of volume-averaged velocity and density is collected on a rectangular grid perpendicular to the flow direction.
Within each grid volume the particle number density and velocities are averaged for $10^3$ timesteps and outputted every $10^4$ timesteps to a file. We found this is a necessary step to avoid excess noise due to particle fluctuations near grid boundaries.
 Training simulations totaling forty output frames were generated. The total time is scaled $1$. In addition, the width of the channel is scaled to be $1$. This scaling is reflected in Figures~\ref{fig:lj_time} thru~\ref{fig:colloid_evo}.
This method allows us to capture initial concentration gradients between species, and their time evolution.
To test the generalization of the operator regression framework, the results given in subsequent sections will be of data that are not directly represented by the training conditions outlined here. In particular, particle densities or colloidal mass ratios that interpolate between training, as well as to times that extrapolate beyond the training set will be used as validation.

We plan to make the configuration files used to run these simulations available at \url{github.com/rgp62/MOR-Physics}.

\subsection{Results} \label{sec:lammps_results}

\subsubsection{Refinement study}

\begin{table}
\begin{center}
    \caption{Hyperparameters for LAMMPS modeling}
    \begin{tabular}{c|c} \label{tab:lammps_param}
        Parameter & Value \\
        \hline
        Network width & 4 \\
        Network depth & 4 \\
        $K$ & 2 \\
        $M$ & 4 \\
        $\beta$ & 1 \\
        Activation function & elu \\ 
        Optimizer & Adam \\
        Learning rate & 0.001 \\
        Batch size & 100 \\
        Epochs & 100
    \end{tabular}
\end{center}
\end{table}

\begin{figure}
\centering
\begin{minipage}{0.4\textwidth}
\centering
\begin{tabular}{c|c} 
Resolution & Error\\ 
\hline
20 & $5.70\times10^{-3}$\\
40 & $5.81\times10^{-3}$\\
80 & $9.85\times10^{-3}$
\end{tabular}
\end{minipage}
\begin{minipage}{2.5in}
\centering
\includegraphics[width=1\textwidth]{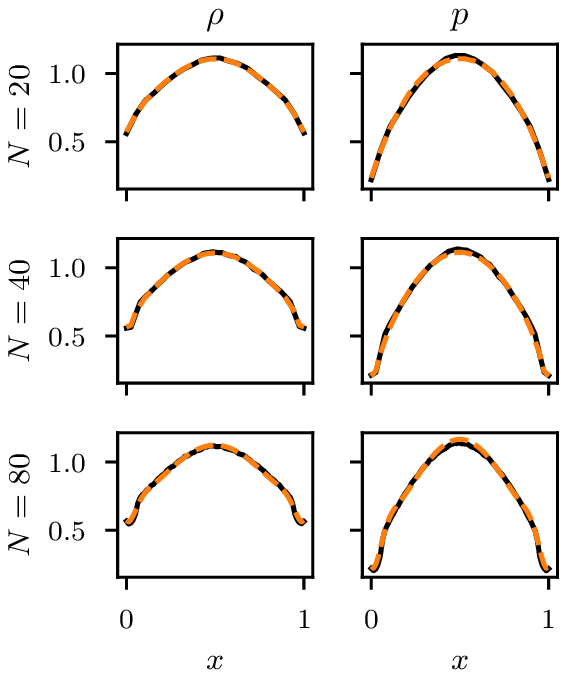}
\end{minipage}
\caption{LJ system with $\rho^*=0.01$. Minimum relative error at steady state over 20 training runs for each grid resolution (\textit{left}). Density and momentum at steady state (\textit{right}) for LAMMPS \sym{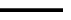} and the learned equation \sym{C1,d,n,1}.}
\label{fig:lj_res}
\end{figure}

For many physical systems, it is difficult to determine the grid resolution needed to accurately capture the dynamics given the ambiguity of defining a representative volume element. Additionally, more resolution may increase the difficulty of training as finer length scale dynamics must be learned. In this study, we examine the effects of resolution on training for the LJ system. We create a dataset with varying grid resolution by performing an ensemble of LJ simulations for $\rho^* = 0.01$ and binning on a fine scale, with $N=80$. The high resolution dataset is obtained by averaging over the ensemble. The lower resolution, $N=40$ and $N=20$ are obtained by downsampling the $N=80$ runs with a flat kernel of width $2$ and $4$ and averaging over one-half and one-fourth of the ensemble, respectively. This ensures the particles sampled per bin is constant among resolutions, controlling for noise. We fit the 2 field model for all three resolutions. Table~\ref{tab:lammps_param} lists the hyperparameters used in this study.

Figure~\ref{fig:lj_res} shows that the extracted models for each resolution agrees well with the LAMMPS simulations at steady state. The $N=80$ case performs worse than the other two cases. As discussed above, higher resolutions may increase the difficulty of training as finer scale dynamics must be learned. In particular, we observe additional flow features in both fields near the walls with increased resolution. The remaining studies use a resolution of $N=20$.

\subsubsection{Model comparison} \label{sec:model}

\begin{figure}[htpb]
    \centering
    \includegraphics[width=4.5in]{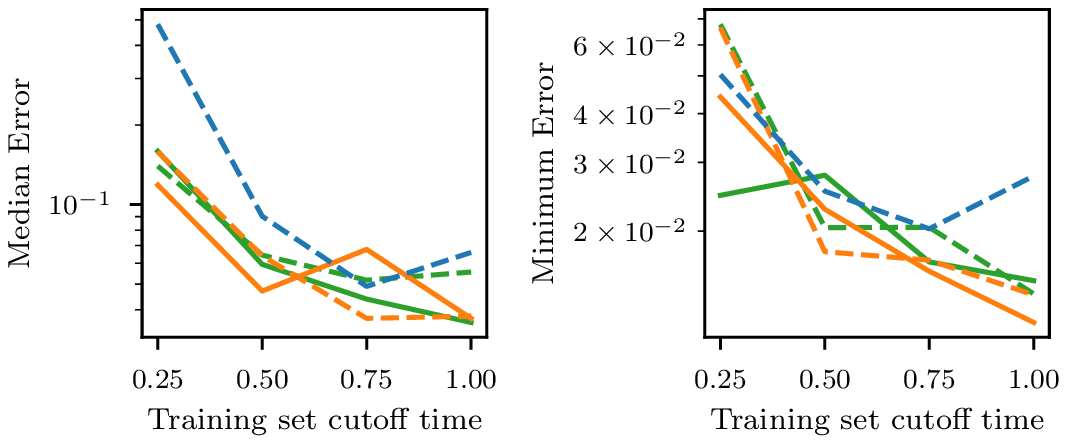}
    \caption{LJ system with $\rho^*=0.01$. Relative error at steady state vs. the cutoff time of training set. Median error (\textit{left}) and minimum error (\textit{right}) over sets of 20 runs is shown for 1 field model \sym{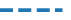}, 2 field model with conservation \sym{C1,l,n,1} and without conservation \sym{C1,d,n,1}, and 3 field model with conservation \sym{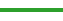} and without conservation \sym{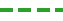}.}
    \label{fig:lj_time}
\end{figure}

\begin{figure}[htpb]
    \centering
    \includegraphics[width=4.5in]{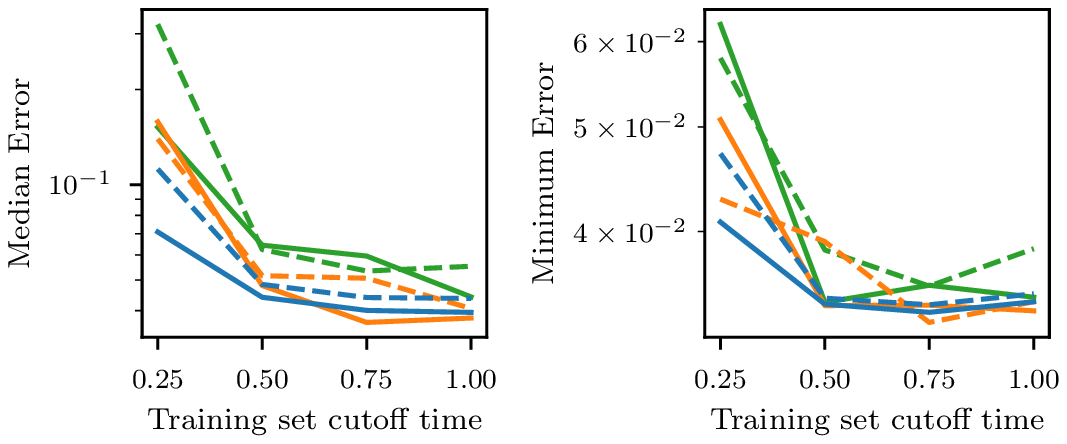}
    \caption{Colloidal system with $(c,d)=(0.2,3)$. Relative error at steady state vs. the cutoff time of training set. Median error (\textit{left}) and minimum error (\textit{right}) over sets of 20 runs is shown for 2 field model with conservation \sym{C0,l,n,1} and without conservation \sym{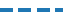}, 4 field model with conservation \sym{C1,l,n,1} and without conservation \sym{C1,d,n,1}, and 6 field model with conservation \sym{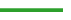} and without conservation \sym{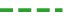}.}
    \label{fig:colloid_time}
\end{figure}

We next compare the models for the LJ and colloidal systems and the benefits of enforcing conservation on $\rho$. We again use the hyperparameters in Table~\ref{tab:lammps_param}. For this study, we vary training data available to the optimizer by including only data up to a cutoff time, $t_c$. For all models, we train over all initial conditions and simulation parameters, i.e., reduced (particle number) density, $\rho^*$, for the LJ system and colloid concentration and particle size, $(c,d)$, for the colloidal system. We test the models' prediction at steady state with the relative error of the momentum for the LJ system and the colloid concentration for the colloidal system.

Figure~\ref{fig:lj_time} compares the 1, 2 and 3 field models for the LJ system. We observe that the model accuracy quickly improves quickly as more of the training data is made available, but quickly levels off. We find that the one equation model performs poorly, while the two and three equation models perform similarly. The one field model, for which only the particle momentum is evolved may not carry enough information for good predictions to be possible. We do not observe a consistent benefit to imposing conservation of $\rho$ in the 2 and 3 field models. The data available may contain enough information for conservation to be inferred or conservation may confer no advantage in the quality of prediction.

Figure~\ref{fig:colloid_time} compares the 2, 4, and 6 field models for the colloidal system. We again observe that the model accuracy is poor for heavily censored datasets but quickly improves and saturates with more data. The 2 and 4 field models perform similarly, while the 6 field model performs poorly. The complexity of the model may have increased the difficulty of training such that proper fits could not be achieved. For this system, we do observe a more consistent, albeit small benefit to imposing conservation of $\rho$.

\subsubsection{Model generalization}

\begin{table}
\begin{center}
    \caption{Hyperparameter optimization ranges}
    \begin{tabular}{c|c} \label{tab:hyp_param}
        Hyperparameter & Range \\
        \hline
        Network width & [1,8] \\
        Network depth & [1,8] \\
        K & [1,4] \\
        M & [1,8] \\
        Learning rate & [1e-5,1e-1] \\
        Batch size & [10,200] \\
        Epochs & [10,1000]
    \end{tabular}
\end{center}
\end{table}

\begin{figure}[htpb]
    \centering
    \includegraphics[width=3.5in]{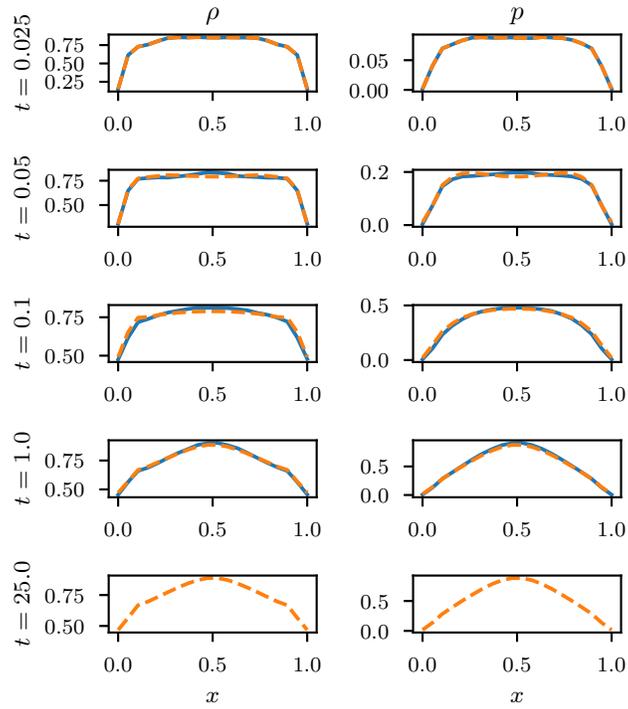}
    \caption{LJ system with $\rho^*=0.1$. Evolution of LAMMPS simulation \sym{C0,l,n,1} and regressed 2 equation model \sym{C1,d,n,1}. Particle density (\textit{first column}), particle momentum (\textit{second column}), and particle temperature (\textit{third column}) is shown for increasing time (\textit{rows}).}
    \label{fig:lj_evo}
\end{figure}

\begin{figure}[htpb]
    \centering
    \includegraphics[width=4.5in]{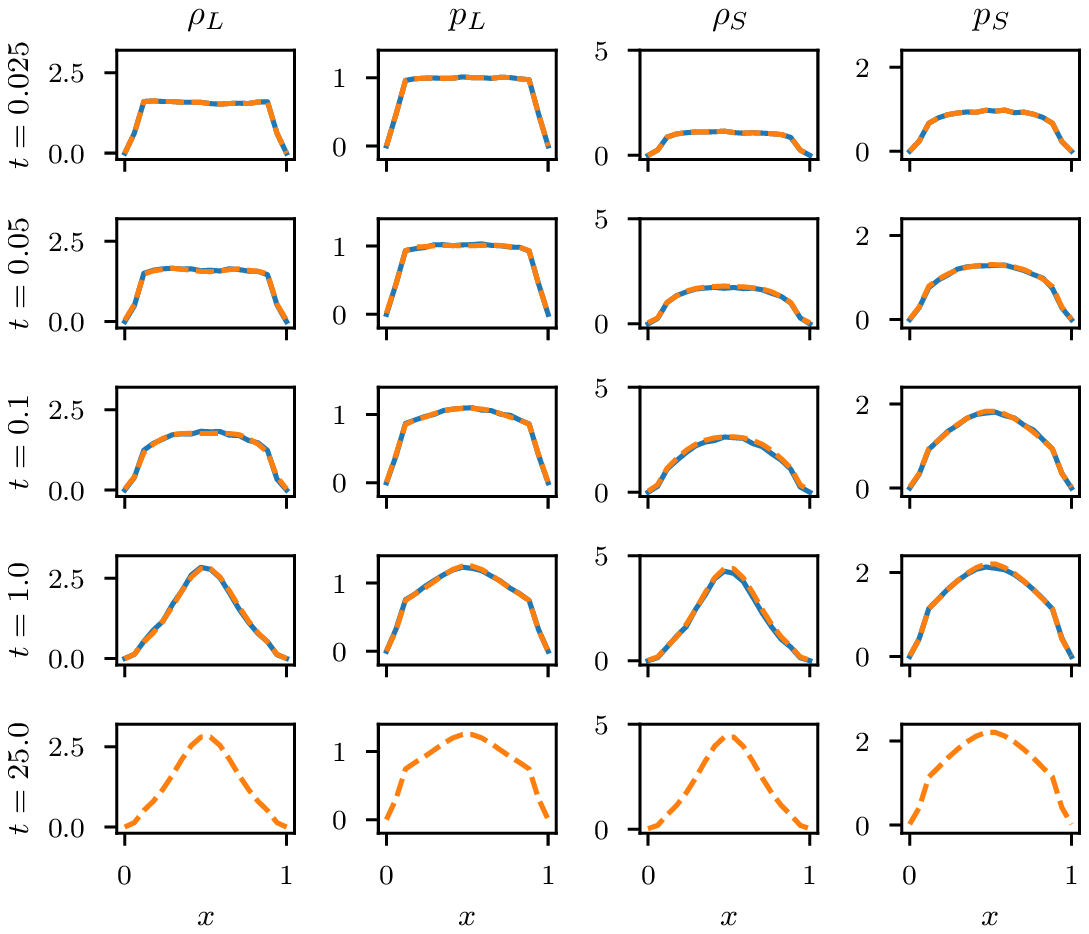}
    \caption{Colloidal system with $(c,d)=(0.15,2.5)$. Evolution of LAMMPS simulation \sym{C0,l,n,1} and regressed 4 equation model \sym{C1,d,n,1}. Small particle density (\textit{first column}), small particle momentum (\textit{second column}), large particle density (\textit{third column}), and large particle momentum (\textit{fourth column}) is shown for increasing time (\textit{rows}).}
    \label{fig:colloid_evo}
\end{figure}

In this section, we attempt to capture the transient behavior of the two systems and evaluate the ability of the models to interpolate between simulation parameters. To perform this study, we separate the LAMMPS data into training sets and test sets. The LAMMPS runs, $\rho^*=0.1$ for the LJ system and $(c,d)=(0.15,2.5)$ for the colloidal system are set aside as the test sets. All other runs form the training sets. We use the 2 field model and 4 field models for the LJ and colloidal systems, respectively.

We found the quality of transient predictions to be heavily sensitive to the choice of hyperparameters. Therefore, we use the black box hyperparameter optimization tool, scikit-optimize \cite{scikitopt}, to tune the models. We use the Bayesian optimization tool within scikit-optimize to identify the optimal set of hyperparameters shown in Table~\ref{tab:hyp_param} with the ranges searched. The default settings for this tool are used. To find the optimal hyperparameters that result in good fits for both the steady state and transient, we use the following loss,
\begin{equation}
    \mathrm{Loss} = \left. \frac{||\bm{u}^{\mathrm{regressed}} - \bm{u}^{\mathrm{LAMMPS}}||}{||\bm{u}^{\mathrm{LAMMPS}}||}\right|_{t=0.025}^2 + \left. \frac{||\bm{u}^{\mathrm{regressed}} - \bm{u}^{\mathrm{LAMMPS}}||}{||\bm{u}^{\mathrm{LAMMPS}}||}\right|_{t=1}^2.
\end{equation}
This loss penalizes hyperparameters that result in poor fits for both the transient and steady state dynamics. The hyperparameter optimization is performed separately for the LJ and colloidal systems.

After training with the optimal hyperparameters, we use the models to predict the evolution of the test systems, shown in Figures~\ref{fig:lj_evo} and \ref{fig:colloid_evo} for the LJ and colloidal systems. For both systems, we observe the models have good agreement with the LAMMPS simulations, up to the final time of the LAMMPS simulations. Additionally, the steady state for the model appears to be stable, even well past the end time of the LAMMPS simulation.

\section{Conclusion}
In this work, we extend a PDE discovery method based upon pseudo-spectral approximation \cite{patel2018nonlinear} to systems of PDEs, multidimensional problems, and domains with non-trivial boundary conditions. Assuming a first order in time PDE, we parameterize the spatial operator as the composition of a Fourier multiplier and a nonlinear point-wise functional, both parameterized by neural networks. Such a paramterization allows for simple introduction of physically motivated inductive biases, such as conservation, translational invariance, isotropy, and reflective symmetry. We demonstrate that we can recover the heat equation and fractional heat equation from density measurements of Brownian motion and Levy flight trajectories in \oneD and \twoD. The inductive biases improve the accuracy of the regression technique and allows it to discover models that extrapolate beyond the dynamics present in the training set. Additionally, we demonstrate that the method can extract suitable continuum models from molecular dynamics simulations of a Lennard-Jones fluid and a colloidal fluid under Poiseuille flow.

Future work will focus on translating this framework to other discretizations, such as the spectral element method, so that complex geometries may be handled more naturally. Additionally, we will focus on applying this method to extract physics in realistic engineering settings. As we demonstrated in this work, we are able to extract accurate models from fine-grained simulations. Using this framework, we may extract models from a broad range of physics, such as turbulence, multiphase flows, and material mechanics. While this work primarily considered molecular dynamics data, one may generalize to non-synthetic experimental data as well.

\section*{Acknowledgement}
Sandia National Laboratories is a multimission laboratory managed and operated by National Technology and Engineering Solutions of Sandia, LLC, a wholly owned subsidiary of Honeywell International, Inc., for the U.S. Department of Energy’s National Nuclear Security Administration under contract DE-NA0003525. This paper describes objective technical results and analysis. Any subjective views or opinions that might be expressed in the paper do not necessarily represent the views of the U.S. Department of Energy or the United States Government.

The work of R. Patel and N. Trask has also been supported by the U.S. Department of Energy, Office of Advanced Scientific Computing Research under the Collaboratory on Mathematics and Physics-Informed Learning Machines for Multiscale and Multiphysics Problems (PhILMs) project. The work of E. Cyr was supported by the U.S. Department of Energy, Office of Advanced Scientific Computing Research under the Early Career Research Program. 
SAND Number: SAND2020-5359 J.

\bibliographystyle{elsarticle-num}
\bibliography{ref}{}

\end{document}